\documentclass[onecolumn,letterpaper,11pt,draftclsnofoot]{IEEEtran}
\usepackage{amsfonts}

\usepackage{amssymb}
\usepackage{amsmath}
\usepackage{bm}
\usepackage{graphicx}
\usepackage{subfigure}
\usepackage{cite}
\usepackage{enumerate}
\usepackage{epstopdf}
\usepackage{multicol}

\usepackage[noend]{algpseudocode}

\usepackage{algorithmicx,algorithm}
\usepackage{hyperref}

\begin{document}

\newtheorem{definition}{\bf Definition}
\newtheorem{theorem}{\bf Theorem}
\newtheorem{lemma}{\bf Lamma}
\newtheorem{proposition}{\bf Proposition}

\title{\huge{ EdgeFlow: Open-Source Multi-layer Data Flow Processing in Edge Computing for 5G and Beyond}}
\author{
\IEEEauthorblockN{
\normalsize{Chao Yao},
\normalsize{Xiaoyang Wang},
\normalsize{Zijie Zheng},
\normalsize{Guangyu Sun},
and
\normalsize{Lingyang Song}
 \\}
\IEEEauthorblockA{\normalsize{School of Electronics Engineering and Computer Science\\ Peking University, Beijing, China\\
Email: \{chao.yao, yaoer, zijie.zheng, gsun, lingyang.song\}@pku.edu.cn} \\
}
}
\maketitle

\thispagestyle{empty}
\begin{abstract}
Edge computing has evolved to be a promising avenue to enhance the system computing capability by offloading processing tasks from the cloud to edge devices.
In this paper, we propose a multi-layer edge computing framework called EdgeFlow.
In this framework, different nodes ranging from edge devices to cloud data centers are categorized into corresponding layers and cooperate together for data processing.
With the help of EdgeFlow, one can balance the trade-off between computing and communication capability so that the tasks are assigned to each layer optimally.
At the same time, resources are carefully allocated throughout the whole network to mitigate performance fluctuation.
The proposed open-source data flow processing framework is implemented on a platform that can emulate various computing nodes in multiple layers and corresponding network connections.
Evaluated on the face recognition scenario, EdgeFlow can significantly reduce task finish time and perform more tolerance to run-time variation, compared with the pure cloud computing, the pure edge computing and Cloudlet.
Potential applications of EdgeFlow, including  network function visualization, Internet of Things, and vehicular networks, are also discussed in the end of this work.
\end{abstract}

\section{Introduction}
As we are moving towards the 5G communication era, various modern applications including Internet-of-Things~(IoT), vehicular networks, mobile caching, and E-health, have been generating tremendous amount of data every day.
The data explosion motivates new challenges and requirements for the equipment upgrade on each device and the computing framework evolution throughout the whole network.
Besides deployment of more powerful servers in cloud data-centers~(CCs), the computation capabilities of wireless access points~(APs), such as macro-cell base stations~(MBSs), small-cell base stations~(SBSs), and WiFi APs, have been improved continuously. In addition, most of them have been equipped with Linux operating systems nowadays~\cite{bib_CRan} to support processing of complex computing programs.
At the same time, the processing power of edge devices~(EDs), such as internet protocol cameras, mobile phones, personal laptops, and smart cars, is also increased rapidly thanks to improvement of System-on-Chip~(SoC) platforms.

Along with the developments of the computing capabilities in the CCs, the APs and the EDs, the computing framework has also evolved as well.
Traditionally, the EDs and the APs are only responsible for data collection and task submission to CCs.
Such a computing model has a number of limitations due to following reasons.
The sustained colossal computation load will incur the enormous resource overload of the CCs, such as on computing resources, energy supply, and cooling systems.
In addition, the remote geographic localization of the CCs will result in the long transmission latency from the EDs to the APs and finally to the CCs, especially when the network includes a great number of EDs and APs and only have limited communication resources.
Therefore, a promising solution, namely \emph{edge computing}, is proposed to leverage the idle computing resources at the edge of the network, i.e., the EDs and the APs, and to save the communication resources as well~\cite{bib_Edge}.\par
\subsection{Existing Edge Computing Platforms}
In an edge computing scenario, part of computing tasks can be offloaded from the cloud end (e.g. the CC), to the edge end
of networks, such as EDs and APs.
When the data have already been processed at the edge end,
%part of the computing tasks can be processed at the edge of the networks, the EDs and the APs.
%After that,
only a small amount of results rather than the raw data in a huge quantity need to be transmitted to the CCs.
Thus, the transmission pressure can be reduced as well \cite{bib_fog3}.\par
Beyond the concept, a number of practical edge computing platforms have already been designed, where some typical ones are listed as follows.
\begin{itemize}
 \item{Cloudlet}: Cloudlet is proposed to reduce the transmission delay through letting the data generators, (usually the EDs), send the computing tasks to the nearest deployed servers rather than the remote CCs, where the WiFi APs are selected to help collect data from the EDs and then send the data to the servers nearby~\cite{bib_cloudlet}.
 \item{Femto Cloud}: Femto Cloud is a fog computing platform that leverages the nearby underutilized EDs, to serve the computing tasks at the network edge, which uses the greedy heuristic optimization model to schedule the incoming tasks~\cite{bib_femto}.
 \item{Paradrop}: Paradrop is an edge-computing platform deployed on the smart WiFi routers~\cite{bib_paradrop}. With a complex computer equipped inside the routers, the Paradrop can enable new applications involving video, e.g. augmented reality, sensor-actuator coordination, and educational applications without the assistance of the remote CCs.
 \item{Iox}: Iox is a fog device product from the Cisco~\cite{bib_iox}. Similar to the Paradrop, the Iox works by hosting applications in a guest operating system running directly on the smart router. It is mainly developed to support ubiquitous IoT business applications.
\end{itemize}
\par
Although existing platforms demonstrate the potentials to implement the edge computing in practical networks, they have some common limitations.
First, most of them only exploit the computing resources in the edge end. For example, Cloudlet, Paradrop and Iox try to leverage
computing power on the APs, while Femto Clould tries to utilize the processing power on the EDs. However, \emph{the coordination of the whole computing resources} throughout the CCs,  the APs and  the EDs, is still not well exploited.
Second, all existing platforms only addressed that processing time can be reduced when the EDs and the APs process more data.
Although it can alleviate the transmission pressure of the data, it may \emph{aggravate the computing pressures} on the EDs and the APs at the edge~\cite{bib_tradeoff}.
How to balance the computing and communication trade-off still remains as an open problem.
Third, existing solutions are normally proposed based on an assumption that the run-time environment is stable. However, in real scenarios, run-time variations, such as data burst in some ED, can have impact on processing efficiency. Thus, it may cause significant performance fluctuation.\par

\subsection{EdgeFlow}
Targeting the issues mentioned above, an EdgeFlow framework is proposed to coordinate the task partitioning among all  data processing devices, and deal with the computing and communication trade-off through the optimal resources allocation.\par
The EdgeFlow is composed of multiple layers. In this work, we categorize all devices into three layers. At the bottom layer, various EDs are located. The data in the EdgeFlow is continuously generated by each ED as a flow. The middle layer includes different types of APs. The top layer is normally a CC. Note that the total system can be further extended to more layers as required in real scenarios. We focus on the three-layer case in this work to simplify discussion.

Each device in the EdgeFlow possesses some computing resources, e.g. CPUs.
When a user submits a data processing task (usually directly notified to the CC), the EdgeFlow can assign part of the task to the EDs,  part to the APs, and the rest in the CC.
The task offloading can directly determine how much computing resources of each device is needed.
When the data has been fully processed at the lower layer, only the results need to be transmitted to the upper layer.
Otherwise, the raw data needs to be transmitted to the upper layer.
Then, the limited communication resources, especially the wireless resources, e.g. time slots, which are supporting the transmission from the lower layer to the upper layer,
can also be optimally allocated in the EdgeFlow.
The algorithm of the entire task division and the computing and communication resources allocation is summarized in a time aligned task offloading~(TATO) scheme.
For implementation, the demo platform is deployed on the Intel Next Units of Computing (Intel NUCs) and the Universal Software Radio Peripherals (USRPs)\cite{ref2}, which is available in  \cite{bib_DemoCode}.
It can emulate various computing nodes in multiple layers and corresponding network connection.
Evaluated on the face recognition scenario, EdgeFlow can significantly reduce task finish time and is tolerant to run-time variation.\par
%The demo is available at.

The rest of the article is organized as follows: The system architecture is given in Section \uppercase\expandafter{\romannumeral2}.
The system schedule is presented in Section \uppercase\expandafter{\romannumeral3}
and the TATO scheme is presented in Section \uppercase\expandafter{\romannumeral4}, respectively.
The demo is implemented in Section \uppercase\expandafter{\romannumeral5}.
The potential applications for EdgeFlow are discussed in Section \uppercase\expandafter{\romannumeral6}.
Finally, the conclusion is given in Section \uppercase\expandafter{\romannumeral7}.

\section{System Architecture}\label{system_model}
The system architecture of EdgeFlow is shown in Fig.~\ref{system_model}.
The bottom layer includes a large number of EDs, such as wireless sensors, mobile phones, and personal laptops.
The middle layer includes the APs,
such as SBSs, MBSs and WiFi APs.
The top layer is a CC, including multiple servers.
Each ED is connected to at most  one AP by wireless links, while each AP is connected to the CC by wired links.
%Nodes on the same layer are not connected directly, which is normally true in real scenarios.
Any one of all the nodes in this architecture is assumed to have a certain amount of computing and communication capabilities.
An online user is able to inquire some information from the system by initiating a task and assigning it to the CC.
The CC is able to complete the task with the help of the APs and the EDs by utilizing their computing and communication abilities.
%Thus, each device undertakes part of the data processing tasks and transmit results and unprocessed raw data to a node on the upper layer.
In the following subsections, the system functions of these three layers are listed in detail.

\begin{figure}[!thp]
\centering
\includegraphics[width=5in]{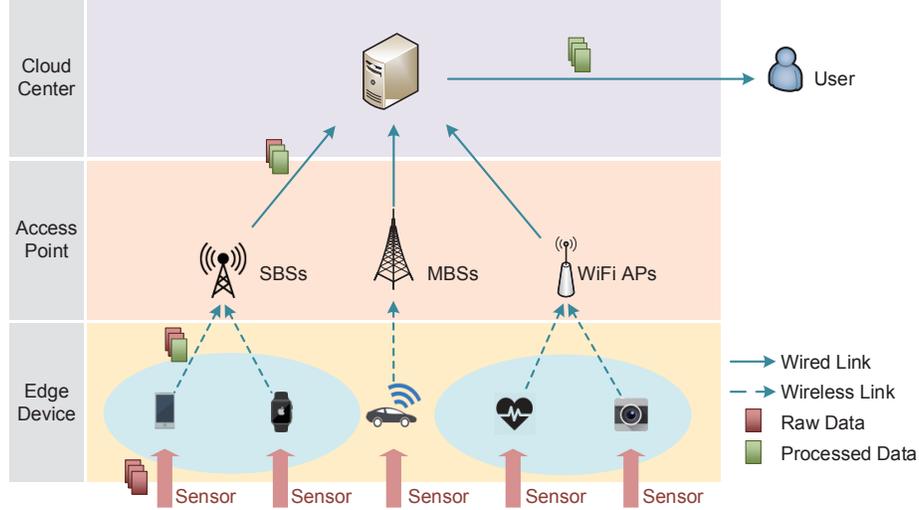}
\caption{The architecture of the EdgeFlow framework.}\label{system_model}
\end{figure}

\subsection{Edge Devices}
Generally, the data flow that related to the users' tasks is generated in the EDs~\cite{bib_dataflow}.
The EDs are responsible for the jobs of sensing, collecting, and generating raw data.
With a certain amount of computing ability, each ED is able to process some of the raw data and submit the unprocessed data and the processed data together to its corresponding AP.
%The EDs first undertake the jobs of sensing, collecting, and generating the raw data. That is to say, the data flow usually starts from the EDs~\cite{bib_dataflow}. Then, the EDs can process part of the computing tasks and use the wireless links to deliver the data to the APs, including the processed data and the rest of the raw data.\par

\subsection{Access Points}
Each AP receives the raw data from its controlled EDs.
Correspondingly, to facilitate the transmission between the AP and the EDs,
the wireless transmission resources allocation among the EDs are also scheduled by the AP.
Besides, similar with each ED, each AP can continue to process a part of the raw data from  EDs, and then use the wired link to submit data to the CC on the top layer. These data include those results processed by the EDs and APs, and the rest of the raw data.\par
\subsection{Cloud Center}
The CC can collect the data from the APs through the wired links and process the rest of the raw data. Then, the CC forwards the final result to the user that generates the task.
Furthermore, the CC gathers the global information and carries out the task offloading strategy, e.g. to decide the amount of data processed at each device for each layer and help calculate the optimal computing and communication resources allocation.\par

\section{System Schedule}
\begin{figure}[!thp]
\centering
\includegraphics[width=5in]{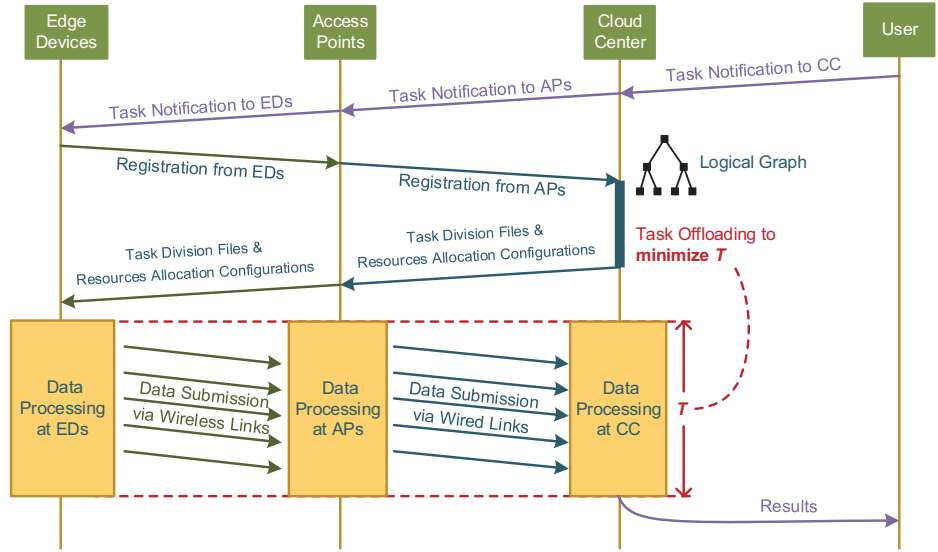}
\caption{The system schedule of EdgeFlow.}\label{uml}
\end{figure}

In this section,
the system schedule of EdgeFlow is presented, as shown in Fig. \ref{uml}.
%shows the system schedule of the EdgeFlow. In the following subsections,
There are four procedures, including task notification, system registration, task offloading, and data processing,
which are shown in detail. \par

% \begin{figure}[!thp]
%\centering
%\Includegraphics[Width=5In]{Schedule.eps}
%\Caption{The Edgeflow System Schedule.}\Label{uml}
%\end{figure}
\subsection{Task Notification}
The user submits a task to the CC. Then, the CC broadcasts the task to the APs. Finally, each AP broadcasts the task notification information to the EDs which it controls.
It is only a short message which notifies the EDs and the APs that the CC is ready to deploy the application.\par
\subsection{System Registration}
After receiving the task notification information,
each device estimates its own computing capability and decides whether to participate in the task (depending on its idle computing resources).
Then, the EDs and the APs upload their registration information with the amount of the available computing and communication resources to the CC. After that,
the CC can create a logical graph of the involved nodes with the information of the resources. \par
\subsection{Task Offloading}
After the CC receives the information from the available EDs and APs, the CC determines a task offloading strategy, TATO, which is introduced in detail in the following Section \uppercase\expandafter{\romannumeral4}.
Based on TATO, the CC assigns the task execution environment, the task division files and the resources allocation configurations to the EDs and the APs.
The task execution environment tells each node how to process the task, which only needs to be assigned once in each task.
The task division file is utilized to tell each device how much data it will process.
The resources allocation configuration tells each device the amount of computing resources.
Besides, the schedule configuration also tells each AP how to allocate the wireless communication resources among the EDs it controls, and how much wired bandwidth it can use for data submission to the CC.\par
\subsection{Data Processing}
After the CC completes the task offloading scheme, the system starts the processing procedure.
The data processing has five stages.
\begin{itemize}
\item{Data Processing at Each ED:} Each ED collects the raw data and processes the part of the data decided by TATO.
\item{Data Submission to Each AP:} Each ED sends the processed results and the rest of the raw data to the corresponding AP through the wireless link.
\item{Data Processing at Each AP:} Each AP processes the part of the data decided by TATO.
\item{Data Submission to the CC:} Each AP delivers its own processed data, the processed data from its controlled EDs, and the rest of the raw data to the CC through the wired link.
\item{Data Processing at the CC:} The CC processes the rest of the raw data. Finally, the CC summarizes the results and submits them to the user.
\end{itemize}

To guarantee the efficiency of the task-offloading strategy,
the EdgeFlow framework will periodically estimate the computing and communication resources.
When the CC detects the significant change of the resource conditions,
the framework will update the task-offloading strategy timely.
\section{Time Aligned Task Offloading}\label{performance_analysis}
Before showing the details of TATO, we formulate the task offloading as a mathematical problem.
Then, we explain TATO for the case with one ED and one AP as well as the case with multiple EDs and multiple APs, respectively. Finally, from the perspective of the generation speed of the data flow, we analyze the properties of TATO.\par
\subsection{Analytical Model}
\begin{figure}[!thp]
\centering
\includegraphics[width=6in]{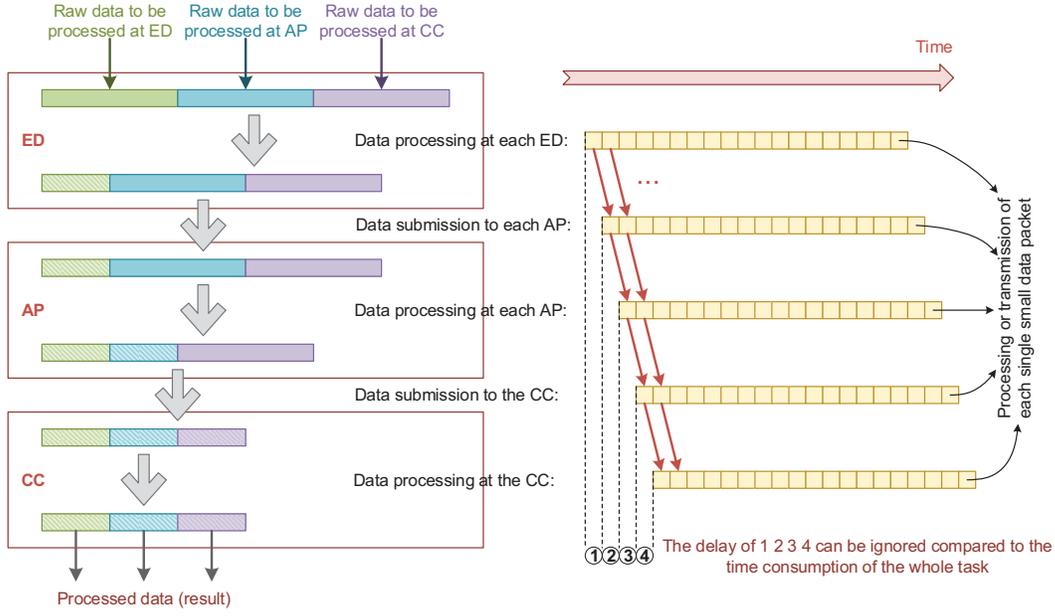}
\caption{The pipeline of the data flow in the EdgeFlow system.
  }\label{divide}
\end{figure}
After one user submits a task, the data are generated at a speed of $\lambda$ on each ED, i.e., a data flow. Then, in any given time span, $\Delta$, the data that each ED can generate are
$\lambda\Delta$. We use the task division percentage parameters $s_{ED}$, $s_{AP}$, and $s_{CC}$ to describe the data that each ED, each AP, and the CC need to process, respectively. Then,
as depicted in Fig.~\ref{divide}, to cope with the data flow at a speed $\lambda$,
there are five stages to process or transmit the data, of which the spent time can be calculated as:
\begin{itemize}
\item{Data Processing Time at Each ED, $C_{b}$:} The amount of the data that each ED needs to process is determined by $s_{ED}$, $\lambda$, and $\Delta$. Then, $C_{b}$ can be calculated as $s_{ED}\lambda\Delta/\theta_{ED}$, where $\theta_{ED}$ is the computing throughput of each ED in one second.
\item{Data Submission Time to Each AP, $D_{b}$:} The data, which an ED needs to transmit to the AP, include the processed data by the ED and the rest of the raw data. Then, $D_{b}$ can be calculated as $(\rho s_{ED}+s_{AP}+s_{CC})\lambda\Delta/\phi_{ED}$, where $\rho$ is the compression ratio after the data processing, and $\phi_{ED}$ is the transmission speed which depends on the wireless communication resources allocated to the ED \cite{bib_model}.
\item{Data Processing Time at Each AP, $C_{m}$:} The raw data arriving at each AP can be calculated as $(s_{AP}+s_{CC})\lambda\Delta$.
Among them,
the amount of the data that each AP needs to process is $s_{AP}\lambda\Delta$.
Then, $C_{m}$ can be calculated as $s_{AP}\lambda\Delta/\theta_{AP}$, where $\theta_{AP}$ is the computing throughput of each AP in one second.
\item{Data Submission Time to the CC, $D_{m}$:} The total data, which an AP needs to submit, include the processed data by the ED, the processed data of its own, and the rest of the raw data. Similar with the analysis of the ED, the data submission time, $D_m$, can be calculated as $(\rho s_{ED}+\rho s_{AP}+s_{CC})\lambda\Delta/\phi_{AP}$, where $\phi_{AP}$ is the transmission speed which depends on the wired bandwidth allocated to the AP.
\item{Data Processing Time at the CC, $C_{t}$:} The raw data arriving at the CC can be calculated as $s_{CC}\lambda\Delta$.
Then, $C_{t}$ can be calculated as $s_{CC}\lambda\Delta/\theta_{CC}$, where $\theta_{CC}$ is the computing throughput of the CC.
\end{itemize}

These stages can be regarded as working concurrently (e.g. when the ED transfers a litter data to the AP,
the AP can start to process the data quickly).
The whole data processing and transmission on the data flow can be regarded as a \emph{pipeline} (where EDs, APs and the CC are the workers  who process their incoming data on an assembly line in order).
% (e.g. when the ED transfers a litter data to the AP, the AP will start to process the data quickly.).
\textbf{The processing or submission time} is the time of the specific layer to process the corresponding data or transmit all the data to the upper layer.
In the pipeline system,
the task finish time depends on the longest time among all pipeline stages mentioned above, $T_{max}$.
When the time of a data processing stage, $C$, equals to $T_{max}$,
it indicates that the computing throughput is the system's bottleneck. When the time of a data transmission stage, $D$, equals to $T_{max}$,
it indicates that the transmission speed is the system's bottleneck, which limits the task finish time.
The objective of TATO is to minimize the longest time $T_{max}$ of the above mentioned five consuming stages.
%%The objective is to minimize the longest time in the data processing and transmission:
%%\begin{equation}\label{formulate}
%%    \min_{\bm{s},\bm{\theta},\bm{\phi}} T_{max}=\min_{\bm{s},\bm{\theta},\bm{\phi}} \max \{ C_{b}, D_{b}, C_{m}, D_{m}, C_{t} \},
%%\end{equation}
%%where vector $\bm{s}$ represents the task division, vector $\bm{\theta}$ represents the computing speeds of all devices, which is determined by the computing resources allocation, and vector $\bm{\phi}$ represents the transmission speeds between pairs of devices, which is determined by the transmission resources allocation. \par

\subsection{TATO with One ED and One AP}
TATO is proposed to help divide the task and allocate the computing and communication resources. The network with \emph{one ED, one AP, and the CC} is used to demonstrate the computing and communication trade-off, the time-aligned principle, and the specific process of TATO.\par
\subsubsection{Computing and Communication Tradeoff}
The computing and communication tradeoff exists on each device. For example, when an ED processes more data, it will consume more computing resources. However, since more data has been processed into the compressed results, the ED will transmit less data to the AP. Similar tradeoff can be observed in the AP. Thus, TATO will first balance the computing and communication tradeoff based on the time calculated in Section \uppercase\expandafter{\romannumeral4}-A.\par
\subsubsection{Time-Aligned Principle}
Since the whole data processing and transmission can be regarded as a pipeline, the ideal case is that all parts in the EdgeFlow keeps working. That is to say, the time of all the data processing and transmission stages remains equal, namely time-aligned principle.
However, this case can hardly happen when we analytically solve the time minimization problem in Section \uppercase\expandafter{\romannumeral4}-A.
The time of some processing stages cannot reach the longest time, which indicates that they work faster than the slowest ones and part of the computing (or communication) resources are wasted. Fortunately, the time-aligned principle can also help to solve the problem even though the most ideal case cannot happen.
In the non-ideal cases,
based on the time-aligned principle,
when we make as many the time periods of the stages equal to the longest time as possible,
the minimization of the time can analytically prove to be solved.
%The minimization of the time can be analytically proved to reach, when we try to make the time of as many as stages keep equal to the longest time.
This principle can help to design the specific process of TATO.\par
\subsubsection{TATO Scheme}
\begin{figure}[!thp]
\centering
\includegraphics[width=6in]{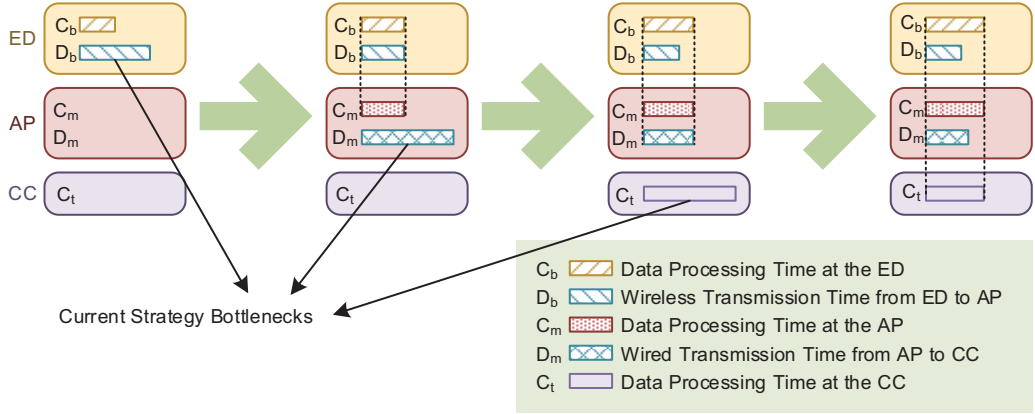}
\caption{TATO for the network with one ED, one AP, and the CC.
  }\label{algorithm1}
\end{figure}

TATO is divided into three steps, which are stated as follows and illustrated in Fig.~\ref{algorithm1}.
\begin{itemize}
\item{Step 1. Task division at the ED:} When $C_{b}>D_{b}$,
it takes more time for the ED to process the data than to transmit the data. This indicates that the ED uses too many computing resources and wastes some transmission resources. Therefore, the data processed by the ED will be reduced and the ED will transmit more raw data.
If $C_{b}<D_{b}$, it takes more time to transmit the data to the AP.
%which causes the computing resources are not taken fully use of on the ED.
This means that the computing resources are not fully used on the ED.
Therefore, TATO will let the ED process more data.
Through this way, the algorithm reaches the optimal result where %$C_{b}=D_{b}=T_{max}^{b} = \max \{C_{b}, D_{b}\}$%
$C_{b}$ and $D_{b}$ achieve the optimal trade-off point at the ED, $T_{max}^{b}$
~\footnote{A special case is that $C_{b}\textgreater$$D_{b}$ still happens even though all data are processed by the ED. This indicates that the transmission is too slow. Thus, the optimal solution is to let all data be processed by the ED.}.\par

\item{Step 2. Task division at the AP:}
To fully use the computing resources at the AP,
we initiate the task division to maximize $C_{m}$ under the limitation that $C_{m}= T_{max}^{b}$.
Then, when $D_{m}\leq T_{max}^b$,
the transmission speed is not the bottleneck.
Hence, the algorithm achieves an optimal solution.
When $D_{m} >T_{max}^{b}$, it takes more time to transmit the data to the CC, which results in the waiting of the data transmission.
Then, the system allocates more data to the ED for processing, and returns to Step 1.
Through iterations~(or analytically solutions), the algorithm reaches the optimal result, where
%$C_b=C_{m}=D_{m}=T_{max}^{m} = \max \{C_{m}, D_{m}, T_{max}^{b}\}$.
$C_{b}$, $C_{m}$ and $D_{m}$ achieve the optimal trade-off point at the AP, $T_{max}^{m}$.

\item{Step 3. Task division at the CC:}
At the CC, all the rest of data should be processed.
When $C_{t}<T_{max}^{m}$, EdgeFlow reaches an optimal solution.
When $C_{t}> T_{max}^{m}$, it takes more time to process the data by the CC. Then, TATO will
process more data at the ED and the AP, and then repeat Step~1 and~2,
to reduce the processing time $C_{t}$.
With iterations~(or analytically solutions), the algorithm reaches the optimal result, where
%%$C_b=C_m=C_t=T_{max}$%%.
$C_{b}$, $C_{m}$, and $C_{t}$ achieve the optimal trade-off point at the CC, $T_{max}$. \par

\end{itemize}
\par
Through the three steps above, the system achieves the optimal solution to minimize the task finish time.\par

\subsection{TATO with Multiple EDs and Multiple APs}
The network with multiple EDs and multiple APs are further considered to demonstrate the resources allocation among devices.
Based on the observations on the case with one ED and one AP, the following corollaries of TATO can be intuitively achieved.\par
\subsubsection{Computing Resources Allocation}
Since the computing resources are held independently from device to device,
each device can solitarily adjust its computing throughput and the computing resources provided for the task.
Observed from the case with one ED and one AP, the optimal point of TATO can be achieved when all devices take full use of their computing resources.
This can also be proved for the cases with multiple EDs and multiple APs.
Then, TATO tries to divide the tasks to let the devices on the same layer have the same data processing time, when all the devices take full use of their computing resources.\par
\subsubsection{Communication Resources Allocation}
For the wireless communication resources, each AP can allocate them on the multiple EDs it controls.
The time-aligned principle also works in this case theoretically.
That is to say, TATO tries to let the transmission time less than or equal to the data processing time on the same devices.
Then, TATO makes as many the time periods of transmission stages equal to $T_{max}$ as possible.
For the wired transmission bandwidth,
%in this version of Edgeflow,
we assume they are independent among different APs.
Then, it is intuitively to take full use of the wired transmission bandwidth for each AP.\par
Similar with the case with one ED and one AP, TATO for multiple devices can also be divided into three steps.
% considering the resources allocation at each step.
Due to the page limit, the similar statements are omitted.\par

\subsection{Performance Analysis on the Data Generation Speed}
Besides the optima on the computing and communication tradeoff, the task division, and the computing and transmission resources allocation, we discuss the properties of TATO under the various data generation speed. \par
\subsubsection{Tasks with Light Data}
The light data indicates that the task finish time is shorter than the data arriving period, $T_{max}<\Delta$. Thus, each device in the EdgeFlow has at least $\Delta-T_{max}$ time to deal with other tasks. Thus, when multiple tasks exist in the network, TATO has the potential to support the multiple tasks when the sum of their task finish time is less than $\Delta$. In this paper, we only consider one task for implementation and the multiple case is left as one of the future directions. \par
\subsubsection{Tasks with Heavy Data}
The heavy data indicates that the task finish time is longer than the data arriving period, $T_{max}>\Delta$.
In other words, a data burst happens.
Thus, the raw data will accumulate on each device. TATO tends to let all the exceed computing time $D-\Delta$ and $C-\Delta$ to be equal on various devices.
which can allocate the overloaded data uniformly on various devices.
The advantage is that when the burst vanishes, the EdgeFlow will process the accumulated data quickly in the parallel manner and recover for the new tasks.\par

%In this section,
%the TATO scheme is proposed to divide the task and allocate the computing and communication resources.
%To fully explore the optimization space, the TATO scheme is designed at a global point of view and works in a centralized manner.
%This is highly relevant to the existing network architecture, where the cloud center controls the system management.
%To achieve more efficient and lightweight method,
%future work should focus on the extension of the proposed TATO scheme on the distributed implementation.

\section{Implementation and Evaluation}\label{simulationresults}

In this section,
we provide the simulations to analyze how EdgeFlow processes the data flow.
These are accomplished by simulating a simple scenario similar to the face recognition application.
%To perform the evaluation,
%we test the TATO scheme and other three heuristic schemes.
In this scenario,
each ED has a camera which collects the image data.
The application aims to recognize the pedestrian faces and slice out the face part, which will be delivered to the CC.
After the analysis of the face part,
the CC will perform the appropriate action.

%\subsection{Performance Analysis}
 \begin{figure}[!thp]
\centering
\includegraphics[width=6in]{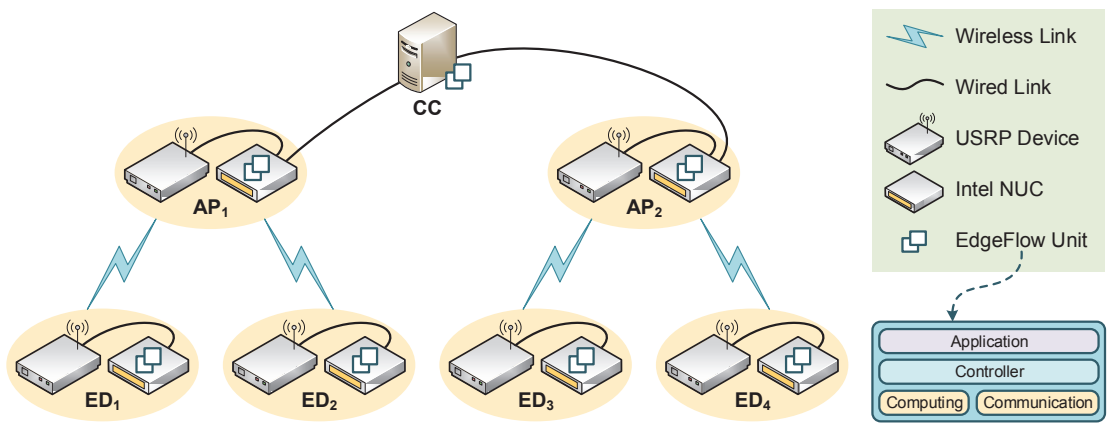}
\caption{The implementation for the EdgeFlow framework.}\label{impla}
\end{figure}

\subsection{Experimental Setup}

As depicted in Fig.~\ref{impla},
there are four EDs, two APs, and one CCs.
One single server stands for the CC layer, and two NUC nodes communicate with it performing as two APs.
The bandwidth of the wired link between the AP and CC is set to $\rm{8}$ Mbps, which is reasonable for the scale of the wireless mesh backbones \cite{bib_parameter}.
To simulate the wireless network with limited resources,
each AP node connects to two ED nodes over USRP devices, which run at the bandwidth of $\rm{5}$ MHz and the transmission power of $\rm{20}$ dBm \cite{ref3}.
%The computing capability is defined as the CPU cycles per unit time.
%%In this application,
%%we use the CPU clock speed to distinguish the computing capabilities among various devices,
%%which means the speed to process the face detection task.
In order to simulate the difference of the computing capabilities among various layers,
the CPU frequencies of each ED, AP and CC are limited to $\rm{1}\times\rm{10^9}$ Hz, $\rm{3.6}\times\rm{10^9}$ Hz and $\rm{36}\times\rm{10^9}$Hz respectively.
By default, the rate of image arrival from the camera is one packet per second, and the average compression ratio after the data
processing is 10\%.

\subsection{Performance Analysis}

To attest the claim that TATO is better than the other three schemes, we make experiments on two scenarios.
The pure cloud computing means the input stream is forwarded to CC directly, and all the processing work is accomplished centrally.
The pure edge computing means each ED deals with all of its input tasks, and delivers the result towards the cloud.
Cloudlet means each ED offloads the face recognition tasks to the corresponding Cloudlet server, which is deployed at the AP.

\begin{figure}[!thp]
\centering
\includegraphics[width=6.5in]{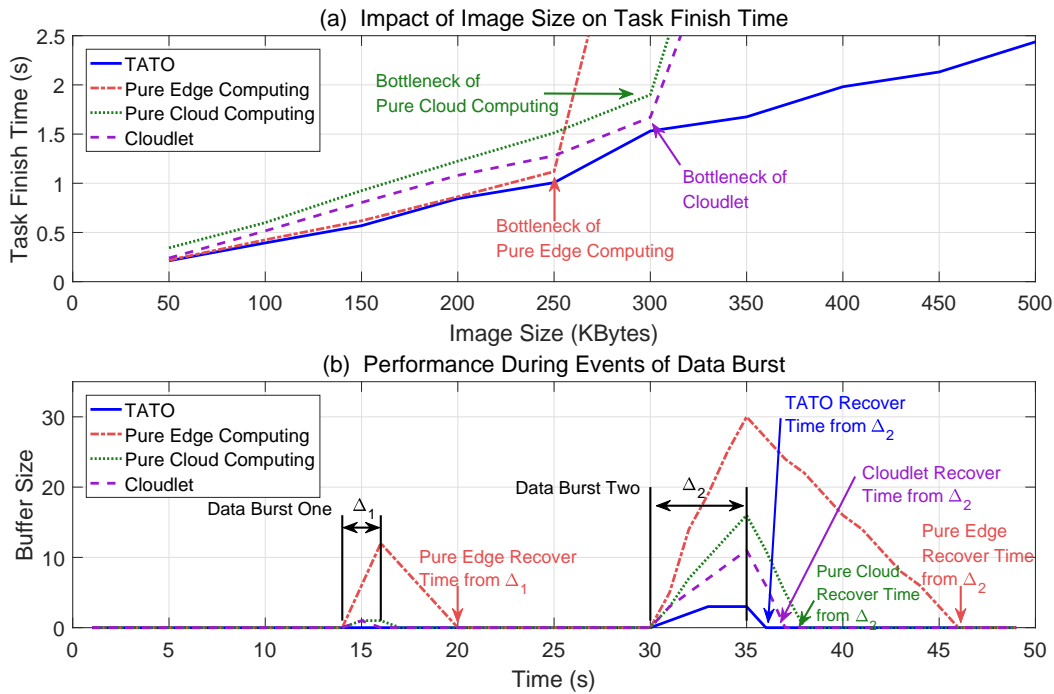}
\caption{Comparison of TATO and the heuristic methods (pure cloud computing, pure edge computing, and Cloudlet) in terms of the image size and system robustness.
  }\label{implement}
\end{figure}

In the first experiment,
we adjust the size of images and observe the resulting average task finish time.
The image size depends on the application monitoring range.
More image data requires more computation and transmission resources.
The task finish time represents the response time from the data generation to the CC performing the appropriate action.
%Since more image size requires more computation and transmission resources,
This experiment studies the efficiency of the schemes under different burdens of tasks.
As shown in Fig.~\ref{implement}(a),
it is plainly evident that TATO is superior in most cases.
%% the TATO performs the best in most cases.
As the size of input data rises, the system may start to run out of resources and unprocessed data start to accumulate. It can be observed that the other three schemes meet their bottleneck earlier than our scheme, with a lower tolerance of data size.

In another experiment,
we analyze system robustness for the tasks with heavy data, which can be reflected by how fast the
buffer size recovers to the stable state after the data burst.
%In the face recognition application,
%when the system wants more detailed
%analysis,
%the EDs will capture more data from the cameras, which may cause the data burst.
The buffer size depends on the sum of the pending images,
which represents the severity of the data burst.
As shown in Fig.~\ref{implement}(b), the first burst causes a data accumulation for the pure edge computing scheme, while the other three are hardly affected.
After that, a bigger burst raises and affects all of the three heuristic schemes.
On the contrary, with the help of TATO, EdgeFlow gains the most robustness for the tasks with heavy data.

Based on the simulations,
when the EDs and the APs possess some communication and computation resources,
TATO could lead to a high system throughput with the coordination of computing resources over all layers
and perform the tolerance to the tasks with heavy data.

\section{Potential Applications}\label{performance_analysis}
%There are huge potentials of offering a range of applications by EdgeFlow.
%By the edge-fog-cloud cooperation, EdgeFlow can provide efficient and reliable computation service to the computation-intensive applications.
%Besides,
%the pre-computing at the edge can shrink the data amount which needs to be transmitted to the cloud center.
%the schedule strategy optimizes the task finish time.
%This drastic shift in data processing paradigm propounded in EdgeFlow can be utilized in many diverse use cases.
In the following,  we introduce three potential applications for the EdgeFlow framework in the 5G communication networks and beyond.
In addition, we clarify the limitations of TATO.

\subsection{Network Function Virtualization}
%Network function virtualization (NFV) is an architecture that uses virtualization technologies to virtualize the system into general-purpose high volume servers.
%Therefore,
%there are abundant free computing and communication resources which can be shared with extra jobs.
%However,
%to utilize the available resources, there are couple of challenges:
%How to ensure the efficiency of the available resources?
%%How to meet the requirements of the computing tasks, such as the task latency?
%%Besides, the cooperation with the classic network architecture should also be carefully designed.
%EdgeFlow provides a ideal choice to leverage the benefits of functions virtualization.
%With the proper task offloading strategy,
%It can achieve the task scheduling with available physical resources,  while will guarantee the system performance.
%%Through the joint task offloading strategy,
%%it can guarantee the system service performance.
%% ensure the optimal utilization of the available resources.
%In addition,
%the system seriously takes the reuse of the existing infrastructure and the difficulty of deployment into account.
%EdgeFlow is compatible with the classic network architecture, without adding any new types of equipment,  which can be easily deployed to the various operating system.
%% in the NFV architecture system.
%%Therefore,
%%this solution can ensure the efficient utilization of the available resources.

Network function virtualization (NFV) is an network architecture which can virtualize the system into general-purpose high volume servers.
Due to the virtualization technology, there are abundant free computing and communication resources on the APs which can carry out extra jobs.
However, how to efficiently utilize the available resources is a significant challenge.
%Furthermore,
%this heterogenous network architecture makes the task offloading problem difficult to analyse.
The EdgeFlow provides an ideal choice to leverage the benefits of functions virtualization.
With TATO,
it can coordinate the task division among all the virtualized APs.
%Besides, the framework design seriously takes into account the reuse of the existing infrastructure and the difficulty of large-scale deployment.
In addition, EdgeFlow is designed based on the open-source Linux operating systems, which can be directly deployed in the NFV architecture, without adding any new type of equipments.

\subsection{Internet of Things}
These are a large scale of IoT sensor applications, whose data can be mined and analyzed,
such as the smart city\cite{ref1}.
%Moreover, the computing task needs to process the data from the multiple sensors as a coherent.
It is evident that the excessive demand for the IoT sensors will quickly overwhelm the  processing speed of the traditional cloud computing architecture.
With the involvement of the EDs directly connected to the sensors and APs, EdgeFlow can enhance the system computing capabilities to meet the explosive data flow in the IoT scenarios.
%Besides, the system is able to schedule the complex operations among the three layers, which is too time-consuming to be computed timely by the IoT sensors.
The data processing before the CC can shrink the amount of the data traffic, which can relieve the communication pressure due to the limited wireless communication resources.
%In summary,
%% some IoT applications need to obtain distributed information for %%computing tasks, such as geo-information, which might be difficult %%for the cloud computing architecture.
%The EdgeFlow system can increase the computation capability for the computation-intensive tasks,
%By the cooperation of the EDs, APs and CC.
%By the pre-computing procedure before the CC, the system can simply and speed up the data analysis while shrinking the amount of the data traffic.
%Moreover, the EdgeFlow system provides the middleware based on different operating systems, which can be convenient for the large-scale deployment.

\subsection{Vehicular Networks}
The vehicular network technology senses the vehicles' behaviors and thus enhances the traffic safety.
The researchers estimate that there are more than one-gigabyte data generated by each car every second.
The data generated from the vehicle require the real-time processing to make the right decision, which severely affect the traffic safety.
Thus, the computing tasks must be offloaded to the vehicles and the roadside units.
%If the computing tasks are all executed in the remote cloud center, it will result in the long task latency.
%Not to mention that there is a huge challenge for the constraint transmission resources to support a number of vehicles at the same time.
The EdgeFlow is able to be an excellent choice for the vehicular networks, which can reduce the response time with the coordination of the computing resources over all layers.
Besides,
the system robustness can handle the traffic congestion scenarios efficiently.
%he system can balance the tradeoff between the computing and communication resources,
%which can enhance the system robustness at the frequent task scenarios with heavy data,
%which is  in the complex traffic scenarios.
%at the data burst scenarios, such as traffic emergency,

\subsection{Limitations and Future work}
However,
EdgeFlow still has some unsuitable scenarios.
The optimal solution with EdgeFlow requires the data to be compressed after the computation procedure.
However,
some applications do not satisfy this requirement (e.g. in some wireless monitoring scenarios, the applications not only analyse the data but also store the historical data).
In these scenarios,
the computation procedure will not significantly decrease the amount of data.
%%This application scenario not only need the computing result but also upload the raw data.
Therefore,
the pre-computing procedure in the EDs and the APs cannot alleviate the transmission pressure.
Future work should analyze how to allocate the task-offloading strategy in these unfavorable scenarios.

%%\section{Limitations and Future Works}

\section{Conclusion}
In this paper, we proposed an open-source multi-layer data flow processing framework, EdgeFlow, which enabled the coordination of the whole computing resources throughout the whole networks.
The task-offloading scheme in EdgeFlow, TATO, can achieve the trade-off between the computing and communication resources and divide the tasks among various layers optimally.
Through the simulations in the face recognition scenario,
TATO can significantly reduce the task finish time and perform a high tolerance to the tasks with heavy data.
The framework has also shown its potential implementations for the 5G communication networks and beyond in some typical applications, such as NFV, IoT, and vehicular networks.
%For the future work, we are going to consider the task-offloading scheme in the unsuitable scenarios, when the pre-computing procedure cannot alleviate the transmission pressure.
%In addition, how to design a lightweight and efficient distributed offloading mechanism would be a key challenge.

%\begin{figure}[!b]
%\centering
%\includegraphics[ width=3.7in]{Pr_FD_big_HD.eps}
%\caption{The probability that FD outperforms HD at the different self interference levels.}\label{fig:FD_big_HD}
%\end{figure}


\begin{thebibliography}{1}

\bibitem{bib_CRan}
A.~Checko, L.~H.~Christiansen, Y.~Yan, L.~Scolari, ``Cloud RAN for Mobile Networks--A Technology Overview," \emph{IEEE Commun. Surveys \& Tutorials}, vol.~17, no.~1, pp.~405-426, Sept.~2014.

\bibitem{bib_Edge}
W.~Shi, J.~Cao, Q.~Zhang, Y.~Li and L.~Xu, ``Edge Computing: Vision and Challenges," \emph{IEEE Internet of Things J.}, vol.~3, no.~5, pp.~637-646, Oct.~2016.

%\bibitem{bib_fog1}
%F.~Bonomi, R.~Milito, J.~Zhu, S.~Addepalli, ``Fog computing and its role in the internet of things," in \emph{Proc. 1st ACM Wksp. Mobile Cloud Computing} ,New York, NY, Aug.~2012, pp.~13-16.
%\emph{ The Workshop on Mobile Big Data}. 2015:37-42.

\bibitem{bib_fog3}
P.~Mach, and Z.~Becvar, ``Mobile Edge Computing: A Survey on Architecture and Computation Offloading," \emph{IEEE Commun. Surveys \& Tutorials}, vol.~19, no.~3, pp.~1628-1656, Mar.~2017.

\bibitem{bib_cloudlet}
M.~Satyanarayanan, P.~Bahl, R.~Caceres, and N.~Davies, ``The Case for VM-Based Cloudlets in Mobile Computing," \emph{IEEE Pervasive Computing}, vol.~8, no.~4, pp.~14-23, Oct.~2009.

\bibitem{bib_femto}
K.~Habak, M.~Ammar, K.~A.~Harras, and E.~Zegura, ``Femto clouds:Leveraging mobile devices to provide cloud service at the edge," in \emph{Proc. IEEE 8th Int. Conf. Cloud Computing,} New York, NY, Aug.~2015, pp.~9-16.


\bibitem{bib_paradrop}
D.~F.~Willis, A.~Dasgupta, and S.~Banerjee, ``Paradrop: a multi-tenant platform for dynamically installed third party services on home gateways," in \emph{Proc. ACM SIGCOMM Wksp.
Distributed Cloud Computing,} Maui, Hawaii, Sept.~2014, pp.~43-44.

\bibitem{bib_iox}
S.~Yi, C.~Li, and Q.~Li, ``A Survey of Fog Computing:Concepts, Applications and Issues," in \emph{Proc. ACM MobiHoc Wksp. Mobile
Big Data}, New York, NY, Jun.~2015, pp.~37-42.

\bibitem{bib_tradeoff}
T.~G.~Rodrigues, K.~Suto, H.~Nishiyama, and N.~Kato, ``Hybrid Method for Minimizing Service Delay in Edge Cloud Computing Through VM Migration and Transmission Power Control," \emph{IEEE Trans. on Comput.}, vol.~66, no.~5, pp.~810-819, Oct.~2017.


\bibitem{ref2}
H.~Zhu, C.~Fang, Y.~Liu, C.~Chen, M.~Li, and X.~S.~Shen, ``You can jam but you cannot hide: defending against jamming attacks for geo-location database driven spectrum sharing," \emph{IEEE J. Sel. Areas Commun.}, vlo.~34, no.~10, pp.~2723-2737, Sept.~2016.

%\bibitem{bib_usrp}
%R.~Dhar, G.~George, A.~Malani, and P.~Steenkiste, ``Supporting Integrated MAC and PHY Software Development for the USRP SDR," in \emph{Proc. 1st IEEE Wksp. %Netw. Technol.
%Softw. Defined Radio Netw.}, Reston, VA, Sept.~2006, pp.~68-77.

\bibitem{bib_DemoCode}
The EdgeFlow framework is available at \url{https://github.com/sirius93123/EdgeFlow}.

\bibitem{bib_dataflow}
I.~Vilajosana, J.~Llosa, B.~Martinez, and M.~Domingo-Prieto, ``Bootstrapping smart cities through a self-sustainable model based on big data flows," \emph{IEEE Commun. Mag.}, vol.~51, no.~6, pp.~128-134, Jun.~2013.





\bibitem{bib_model}
K.~Zhang, Y.~Mao, S.~Leng, Q.~Zhao, L.~Li, X.~Peng, L.~Pan, S.~Maharjan, and Y.~Zhang, ``Energy-Efficient Offloading for Mobile Edge Computing in 5G Heterogeneous Networks," \emph{IEEE Access}, vol.~4, no.~99, pp.~5896-5907, Aug.~2017.



%\bibitem{bib_NFV}
%R.~Vilalta, A.~Mayoral, D.~Pubill, R.~Casellas, R.~Mart¨ªnez, J.~Serra, C.~Verikoukis, and R.~Munoz, ``End-to-End SDN orchestration of IoT
%services using an SDN/NFV-enabled edge node," in \emph{Proc. Optical Fiber Conf.,} Anaheim, CA, Aug.~2016.


\bibitem{bib_parameter}
N.~Kato, Z.~M.~Fadlullah, B.~Mao, F.~Tang, O.~Akashi, T.~Inoue, and K.~Mizutani, ``The Deep Learning Vision for Heterogeneous Network Traffic Control: Proposal, Challenges, and Future Perspective," \emph{IEEE Wireless Commun.}, vol.~24, no.~3, pp.~146-153, Dec.~2016.

\bibitem{ref3}
S.~Fang, Y.~Liu, W.~Shen, H.~Zhu, and T.~Wang, ``Virtual multipath attack and defense for location distinction in wireless networks", \emph{IEEE Trans. on Mobile Computing}, vol.~16, no.~2, pp.~566-580, Feb.~2017.

\bibitem{ref1}
P. L. Lau, N. Wijerathne, B. K. K. Ng, C. Yuen, "Sensor Fusion for Public Space Utilization Monitoring in a Smart City", IEEE J. on Internet of Things, August 2017.



%\bibitem{bib_vehicle}
%H.~Wu, R.~Fujimoto, and G.~Riley, ``Analytical Models for Information Propagation in Vehicle-to-Vehicle Networks," in \emph{Proc. 60th IEEE
%Vehic. Tech. Conf.,} Los Angeles, CA, Apr.~2004, pp.~4548-4552.



%\bibitem{interference_cancel} T. Riihonen, S. Werner, and R. Wichman,
%`` Hybrid full-duplex/half-duplex relaying with transmit power adaptation,''
%\emph{IEEE Trans. Wireless Commun.}, vol. 10, no. 9, pp. 3074--3085, Sep. 2011.
%\bibitem{RiihonenTSP} T. Riihonen, S. Werner, and R. Wichman,
%``Mitigation of loopback self-interference in full-duplex MIMO relays,''
%\emph{IEEE Trans. Signal Process.}, vol. 59, pp. 5983--5993, Dec. 2011.
%
%\bibitem{powercontrol} W. Cheng, X. Zhang, and H.Zhang,
%``Optimal dynamic power control for full-duplex bidirectional-channel based wireless networks,''
%in \emph{2013 Proc. IEEE INFOCOM}, pp. 3120--3128, Apr. 2013.






%\bibitem{Rapacity}H. A. Suraweera, P. J. Smith, and M. Shafi,
%``Capacity limits and performance analysis of cognitive radio with imperfect channel knowledge,''
%\emph{in Proc. IEEE INFORCOM 2013 }, pp. 3120--3128, 2013.
%
%\bibitem{FDcapacity}W. Cheng, X. Zhang and H. Zhang,
%``Optimal dynamic power control for full-duplex bidirectional-channel based wireless networks,''
%\emph{IEEE Trans. Veh. Technol.}, vol. 59, pp. 1811--822, May. 2010.
%
%\bibitem{Abramovitz} M. Abramowitz and I. A. Stegun,
%\emph{Handbook of mathematical functions with formulas, graphs, and mathematical tables}, 9th Edition, NewYork: Dover, 1970.

%\bibitem{Gradshteyn94} I. S. Gradshteyn and I. M. Ryzhik,
%\emph{Table of integals, series, and products}, 5th Edition, Academic Press, 1994.

%\bibitem{wireless Communication} D.Tse and P.Viswanath,
%\emph{Fundamentals of Wireless Communication},Cambridge university press, 2005.

\end{thebibliography}
\end{document}